\documentclass[sigconf]{acmart}

\usepackage{multirow}
\usepackage{dirtytalk}
\usepackage{colortbl}
\usepackage{xcolor}
\usepackage{fontawesome}
\usepackage{soul}
\usepackage{acmart-taps}

\definecolor{new-green}{rgb}{0.104,0.667,0.229}

\newcommand{\rev}[1]{{\color{black}{#1}}}

\newcommand{\textft}[1]{{\fontfamily{lmss}\selectfont{#1}}}

\newcommand{\customul}[2][black]{\setulcolor{#1}\ul{#2}}

\definecolor{naimul-orange}{rgb}{0.9450, 0.5647, 0.1254}

\definecolor{naimul-green}{rgb}{0.4078, 0.6392, 0.4274}

\definecolor{naimul-blue}{rgb}{0.21569,  0.49412,  0.72157}

\definecolor{naimul-purple}{rgb}{0.59608,  0.30588,  0.63922}

\newcommand{\aiwritten}[1]{{\textft{\customul[naimul-orange]{#1}}}}

\newcommand{\aiinfluenced}[1]{{\textft{\customul[naimul-green]{#1}}}}

\newcommand{\promptedit}[1]{{\textft{\customul[naimul-blue]{#1}}}}

\newcommand{\promptgenerate}[1]{{\textft{\customul[naimul-purple]{#1}}}}

\copyrightyear{2024} 
\acmYear{2024} 
\setcopyright{rightsretained} 
\acmConference[CHI '24]{Proceedings of the CHI Conference on Human Factors in Computing Systems}{May 11--16, 2024}{Honolulu, HI, USA}
\acmBooktitle{Proceedings of the CHI Conference on Human Factors in Computing Systems (CHI '24), May 11--16, 2024, Honolulu, HI, USA}
\acmDOI{10.1145/3613904.3641895}
\acmISBN{979-8-4007-0330-0/24/05}

\begin{document}

\newcommand{\toolname}{\textsc{HaLLMark}}


\title[The HaLLMark Effect]{The \toolname{} Effect: Supporting Provenance and Transparent Use of Large Language Models in Writing with Interactive Visualization}

\author{Md Naimul Hoque}
\email{nhoque@umd.edu}
\affiliation{%
  \institution{University of Maryland}
  \city{College Park}
  \state{MD}
  \country{USA}
}

\author{Tasfia Mashiat}
\email{tmashiat@gmu.edu}
\affiliation{%
  \institution{George Mason University}
  \city{Fairfax}
  \state{VA}
  \country{USA}
}

\author{Bhavya Ghai}
\authornote{Bhavya Ghai contributed to this work while he was with Stony Brook University.}
\email{bhavyaghai@gmail.com}
\affiliation{%
  \institution{Amazon}
  \city{New York}
  \state{NY}
  \country{USA}
}

\author{Cecilia Shelton}
\email{sheltonc@umd.edu}
\affiliation{
    \institution{University of Maryland}
    \city{College Park}
    \state{MD}
    \country{USA}
}

\author{Fanny Chevalier}
\email{fanny@dgp.toronto.edu}
\affiliation{
    \institution{University of Toronto}
    \city{Toronto}
    \state{ON}
    \country{Canada}
}

\author{Kari Kraus}
\email{kkraus@umd.edu}
\affiliation{
    \institution{University of Maryland}
    \city{College Park}
    \state{MD}
    \country{USA}
}

\author{Niklas Elmqvist}
\email{elm@cs.au.dk}
\affiliation{%
  \institution{Aarhus University}
  \city{Aarhus}
  \country{Denmark}
  \orcid{0000-0001-5805-5301}
}

\renewcommand{\shortauthors}{Hoque et al.}

\begin{abstract}
    The use of Large Language Models (LLMs) for writing has sparked controversy both among readers and writers.
    On one hand, writers are concerned that LLMs will deprive them of agency and ownership, and readers are concerned about spending their time on text generated by soulless machines.
    \rev{On the other hand, AI-assistance can improve writing as long as writers can conform to publisher policies, and as long as readers can 
    be assured that a text has been verified by a human.}
    We argue that a system that captures the provenance of interaction with an LLM can help writers retain their agency, conform to policies, and communicate their use of AI to publishers and readers transparently.
    Thus we propose \toolname, \rev{a tool for visualizing the writer's interaction with the LLM.}
    We evaluated \toolname{} with 13 creative writers, and found that it helped them retain a sense of control and ownership of the text.
\end{abstract}



\begin{teaserfigure}
\centering
  \includegraphics[width=0.85\textwidth]{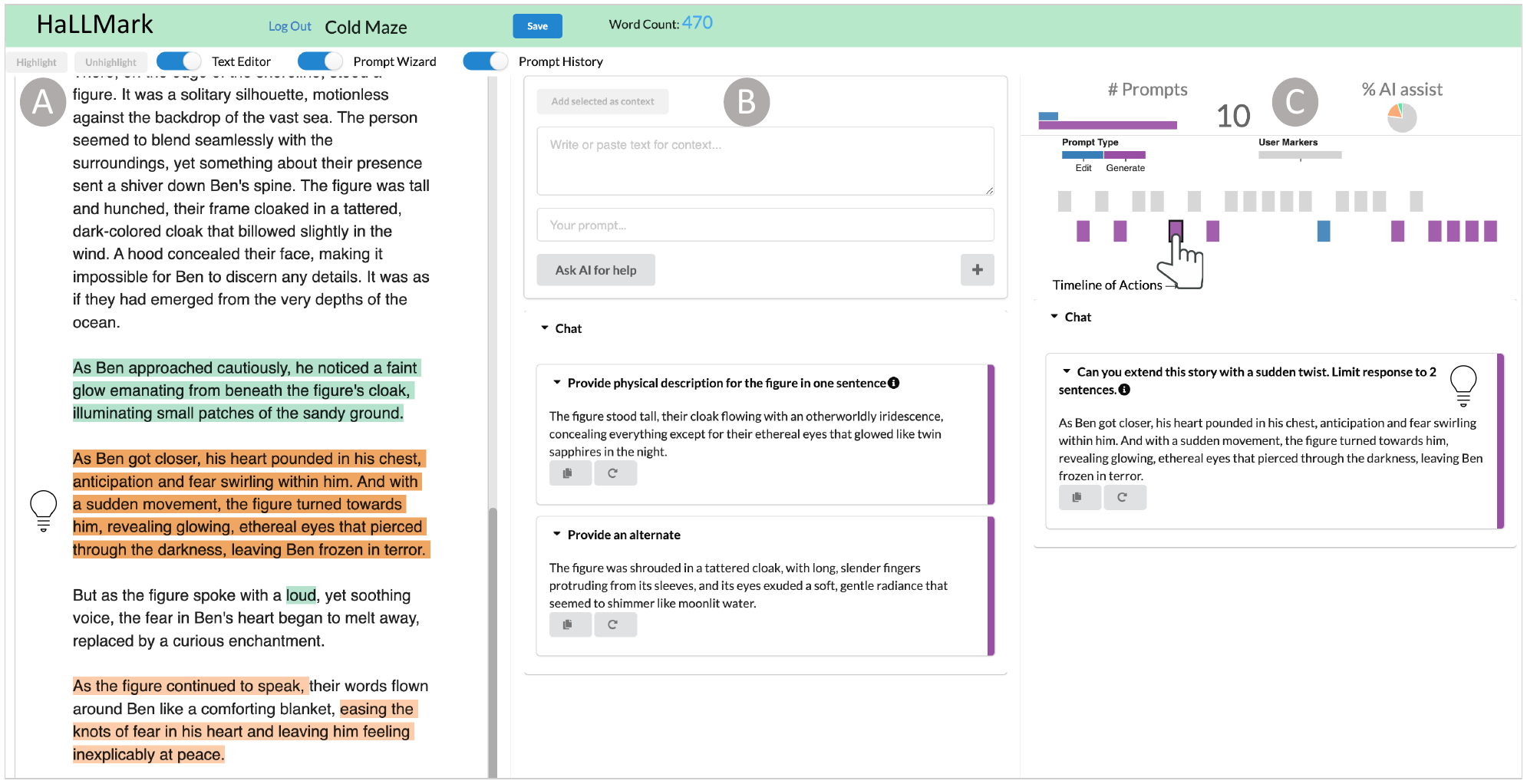}
  \caption{\textbf{The \toolname{} System.}
  \rev{(A) Text editor for viewing and editing text.}
  The system highlights text written (orange) and influenced (green) by the AI. \rev{There are three toggle buttons on top of the editor to turn on and off the three views (columns) of the interface.}
  (B) Prompting interface for large language models (e.g., GPT-4).
  The user can see the prompts and AI responses for the current session.
  (C) Summary statistics show the number of prompts and percentage of user-written text and AI assistance.
  Below that, we see a timeline of a user's writing actions (grey rectangles) and interaction with the AI (purple and blue rectangles).
  The user can hover over any glyphs in the timeline to see the relevant prompt and linked text in the text editor.}
  \label{fig:teaser}
  \Description{The Figure shows a web-based interface, where on the left side it has a text editor, in the middle some input boxes, and on the right there are some visual components.}
\end{teaserfigure}

\maketitle

\section{Introduction}
\label{sec:intro}

Large Language Models (LLMs) are here to stay.
Their arrival has been particularly widely panned in the area of creative writing, where doomsayers are hinting at a future where it will become impossible to distinguish between a writer's original work and text generated by an LLM.
Worse yet, LLMs are feared to potentially inundate us with a cornucopia of poor text.
What happens to original thought if we are subjected to writing forever recycled and regurgitated from original work that has already been produced by our forebears?
It has even been argued that some essential aspect of what it means to be human is lost if creative writing is performed by machines.
While these are valid concerns, it remains that LLMs are just another tool in a long line of tools, and we should find ways to harness the technology in a just, responsible, and ethical way rather than attempt to suppress it.
But in trying to embrace this technology, one critical concern for writers who want to use LLMs is properly attributing contributions from the AI.
This gives rise to an ``ownership tension'' among writers and hampers their agency and control over the process~\cite{DBLP:conf/ACMdis/GeroLC22, DBLP:conf/chi/MirowskiMPE23}.
Identifying contributions from AI is also important for writers who need to conform to AI-assisted writing policies~\cite{copyright, authors_guild}.
Authors currently have few mechanisms to track their accountability with regard to these rules and policies. 

In this work, we argue that capturing interactions between AI and writers as the document evolves (i.e., provenance information) and supporting interactive exploration of such provenance information will \textbf{improve} the writer's agency, control, and ownership of the final artifact (e.g., short stories, novels, poems).
Prior research suggests that externalizing provenance can help balance automation and agency in human-AI collaboration~\cite{Shneiderman2022}.
Provenance can also help writers conform to AI-assisted writing policies and provide transparency to publishers and readers.
To explore this design space, we first reviewed existing guidelines and policy documents on the use of LLMs from several professional, educational, and academic organizations (Section~\ref{sec:formative}).
This review informed us about the types of information that writers should be aware of in AI-assisted writing.
We then developed \textsc{\toolname}, a web-based \rev{technology probe~\cite{10.1145/642611.642616}} that integrates an authoring interface with LLM support that stores and visualizes a writer's interaction with an LLM (Section~\ref{sec:system}).
The system facilitates writers in self-reflecting on their use of AI by clearly highlighting AI contributions.

To validate \toolname{}, we engaged a group of creative writers to use it to write a short story during remote evaluation sessions and collected their feedback and resulting stories (Section~\ref{sec:eval}).
Our findings suggest \toolname{} encouraged writers to actively evaluate AI-assistance from the onset of the writing process.
As a result, it instilled a sense of control in the writer's mind and improved ownership of the final artifact.
Our findings also suggest \toolname{} will help writers conform to AI-assisted writing policies without the need for manually generating disclosures.
Writers were therefore confident that \toolname{} will help them become more transparent and that it is an effective medium to communicate the use of AI in writing to external parties (i.e., publishers).
We close the paper by outlining \rev{how other stakeholders (e.g., readers, peer-reviewers, publishers) could potentially use and benefit from our approach (Section~\ref{sec:disc}).
We also discuss the broader impact of our work on the ethical use of LLMs in the writing and publication industry.}

\section{Related Work}
\label{sec:related-work}

Our work intersects with prior art on intelligent writing support tools, concerns around the use of LLMs, and the use of visualization in literature and writing support tools.
Here we describe each of these topics in turn.

\subsection{Writing Support Tools}

Computers have long been used as writing support tools, \rev{harking back} to the spelling check feature on Microsoft Word in 1997 and all the way to modern-day Large Language Models (LLMs).
In fact, one could argue that the ubiquitous typewriter and printing press are examples of enabling technology for writers.
One popular category of writing support tools is paid software such as Scrivener~\cite{scrivener} and Granthika~\cite{granthika} that try to enhance the organizational capabilities of a writer.
An additional category of writing support tools targets academic writers who need support of various kinds.
The tool with the most relevance to this study is Turnitin~\cite{turnitin}, which is similarly postured to advocate for transparency in original authorship through academic integrity and proper attribution techniques such as citation.
More relevant to our work are tools that enhance creativity of writers through \textit{interactive} and \textit{intelligent} features.
Examples of such works include support for metaphor creation~\cite{DBLP:conf/chi/GeroC19, DBLP:conf/ACMdis/KimSCX23}, automatic text summarization~\cite{DBLP:conf/uist/DangBLB22}, interaction with literary styles~\cite{sterman2020interacting}, and support for the iterative revising process~\cite{du-etal-2022-read, du-etal-2022-understanding-iterative}.

More recently, the introduction of LLMs has fueled a new generation of writing support and \textit{co-writing} tools.
These tools can generate human-like text and inspire new narrative angles and ideas. 
For example, CoAuthor~\cite{lee2022coauthor} and Wordcraft~\cite{yuan2022wordcraft} can generate new sentences and passages to help writers develop short stories. 
Dramatron~\cite{DBLP:conf/chi/MirowskiMPE23} is a similar kind of system but generalizes to long-form writing through hierarchical chaining of prompts. 
Sparks~\cite{DBLP:conf/ACMdis/GeroLC22} focuses on science writing, while TaleBrush~\cite{talebrush} can generate texts to match a character arc sketched by the author.
\toolname{}, the tool proposed in this paper, is built on similar mechanisms but has a different focus.
More specifically, the focus of \toolname{} is to leverage interactive provenance to help writers reflect on their use of the LLM, to conform to new policies on AI-assisted writing, and to retain their ownership as well as transparently communicate the influence of the AI on the text. 

\subsection{Concerns around LLMs as Writing Assistants}

While the capabilities of LLMs---similar to other generative AI tools---have awed writers from different domains, their use for creative purposes is controversial~\cite{Epstein_2023}.
\rev{The Writers Guild of America (WGA) and the Screen Actors Guild -- American Federation of Television and Radio Artists (SAG-AFTRA) were recently on  strike, the former from May to September 2023, and the latter from July to November 2023.}
Along with typical demands such as better pay structure, especially for streaming services, the main demands from protesters \rev{were} to add contract language that protects them from being replaced by machines (writers from AI-generated text \rev{and} actors from studios using their AI-generated likenesses).
Similar sentiments have been reported in several recent studies.
Writer concerns for LLMs include agency and ownership~\cite{DBLP:conf/chi/MirowskiMPE23, DBLP:journals/corr/abs-2112-04359}, ethics and plagiarism~\cite{DBLP:journals/corr/abs-2112-04359}, and lackluster, stereotyped text~\cite{DBLP:conf/chi/MirowskiMPE23, DBLP:conf/ACMdis/GeroLC22}.

Creative writers who are pushing the envelope of technology view generative AI as just another tool that can help them support their work more efficiently.
Even among these supporters, a majority prefer to limit the use of generative AI to supporting their editing, brainstorming, or organizing rather than asking it to creatively generate the text of their work in order to retain their own agency, ownership, and artistic expression~\cite{ai_survey}.
This echoes principles of meaningful human control in generative AI articulated by Epstein et al.~\cite{Epstein_2023}.
Given the paired, if sometimes conflicting, interests of writers who want to both embrace the affordances of new generative AI technologies and also carefully and thoughtfully limit the ratio of AI-generated text output in their final work, it is clear that we need guidelines and policies for ensuring responsible use of LLMs in writing that account for the strengths and weaknesses of the tools as well as ethical concerns regarding their use.

In response, organizations and publishers such as the U.S.\ Copyright Office~\cite{copyright}, Author's Guild~\cite{authors_guild}, and ACM~\cite{acm} have released guidelines for AI-assisted writing.
These policies ask writers to track their interactions with LLMs (i.e., provenance) and report interactions to show that writers had creative control over the generation of the text.
However, it is not clear how writers can operationalize these guidelines in their writing and report use of AI transparently.
This paper first formalizes the guidelines from existing policies into actionable items and then presents a tool that writers can use to conform to the policies.
The result is a tool that supports provenance for authors, helping them regain agency and authorship, while at the same time, allowing them to conform to the policies and transparently communicate the process to others (e.g., readership, publishers).

\subsection{Visualization for Text and Writing}

Data visualization can be particularly helpful for summarizing and representing large volumes of text~\cite{Brath2021}.
Text is largely an unstructured data format, making it difficult to see hidden patterns in text-based artifacts, such as a novel, document collection, or newspaper article.
Text visualization is the area of visualization research that invents new representations to summarize and comprehend text data~\cite{alencar2012seeing, Brath2021}.
Examples of text visualizations include the ubiquitous word cloud~\cite{DBLP:journals/interactions/ViegasW08}, wordle~\cite{DBLP:journals/tvcg/ViegasWF09}, and the Word Tree~\cite{DBLP:journals/tvcg/WattenbergV08}.
More complex representations also convey structure in a text corpus, such as Phrase Nets~\cite{DBLP:journals/tvcg/HamWV09}, TextFlow~\cite{DBLP:journals/tvcg/CuiLTSSGQT11}, and ThemeDelta~\cite{DBLP:journals/tvcg/GadJGEEHR15}.
J{\"a}nicke et al.~\cite{DBLP:conf/vissym/JanickeFCS15} provide a survey on the use of text visualization and text analytics in support of close and distant reading in the digital humanities.

Despite the prevalence of text visualization in the academic community, application of these representations in writing support tools is limited.
One example is DramatVis Personae (DVP)~\cite{DBLP:conf/ACMdis/HoqueGE22}, a visualization system for mitigating nuanced social biases in creative writing.
In follow-up work, the DVP authors developed a tool to visualize different character traits~\cite{DBLP:conf/ACMdis/HoqueGE23}. 
Finally, Poemage~\cite{DBLP:journals/tvcg/McCurdyLCM16} helps literary scholars understand the sonic properties of a poem.

Visualization has long been used to explain machine learning~\cite{DBLP:journals/tvcg/WongsuphasawatS18} and NLP models~\cite{DBLP:conf/avi/ChuangMH12, DBLP:journals/ijmms/LeeSSEBF17}, and these ideas have also recently begun to be applied to LLMs.
Recently, Jiang et al.~\cite{jiang2023graphologue} proposed Graphologue, that converts responses from LLMs to interactive graphs for fast and non-linear sensemaking. 
Sensescape~\cite{suh2023sensecape} supports multilevel organization of information gathered from LLM responses.
However, none of these tools focus on supporting writing, provenance, or transparency\rev{, a gap which our work aims to address}.

\subsection{Visualizing and Tracking Contributions in Collaborative Work}

\noindent\rev{A field of research relevant to our work is visualization methods and systems proposed to track contributions from multiple parties in a collaborative setting~\cite{DBLP:conf/citi/Vasquez-Bermudez21, DBLP:conf/visualization/LiB0L20}.
For instance, researchers have shown that visualization can track provenance in collaborative writing~\cite{10.1145/2145204.2145325, 10.1145/3301275.3302328, 10.1145/2998181.2998356, chevalier2010using}.
History Flow~\cite{DBLP:conf/chi/ViegasWD04} and DocuViz~\cite{DBLP:conf/chi/WangOZNO15} are examples of visualization techniques for studying co-authorship patterns (cooperation and conflicts) in collaborative writing.}

\rev{Several visualization systems have been proposed to track provenance in human-AI collaboration~\cite{7192714}.
For example, tracking model and data performance is a key need for interactively developing machine learning models. Amershi et al. proposed ModelTracker~\cite{10.1145/2702123.2702509} to support this need. ModelTracker is an interactive visualization that summarizes traditional summary statistics and graphs while displaying example-level performance to enable direct error examination and debugging. Chameleon~\cite{10.1145/3313831.3376177} uses a collection of visualizations to allow users to compare data features, splits, and performance across data versions. Other works in this area have focused on visualizing contributions from AI and humans for a specific task. For example, Rogers and Crisan recently proposed AutoML Trace~\cite{10.1145/3544548.3580819}, a system for visualizing contributions from humans and AI in AutoML.
Wu et al.~\cite{10.1145/3491102.3517582} showed that decomposing an LLM task into multiple sub-tasks, chaining them, and then allowing users to investigate how a previous task influences the subsequent task improved users' task quality, sense of control, and system transparency.}

\rev{While these works inspired us to utilize visualization in externalizing provenance information, AI-assisted writing---especially with LLMs---is a fairly new research area with emerging challenges for tracking provenance.
These challenges include understanding the types of information that writers should be aware of in AI-assisted writing, coupling requirements from writers and policies on AI-assisted writing, and supporting writers' needs with interactive visualization.
This paper aims to address these challenges.}

\begin{table*}[t]
    \centering
    \begin{tabular}{l|p{4.5cm}|p{6cm}}
    \toprule
    \rowcolor{gray!10} 
    \textsc{\textbf{Dimension}} & \textsc{\textbf{Category}} & \textsc{\textbf{Examples}} \\
    \hline
     
    \multirow[t]{2}{*}{D1. Prompt category} &  Asking for editing an existing text &  No need to report common editorial assistance (grammar check and paraphrasing)~\cite{acm, acl, copyright}   \\
    \cline{2-3}
    &  Asking for generating new texts & Report prompts if they are used to generate an extensive amount of text~\cite{acm}\\
    \hline

    \multirow[t]{3}{*}{D2. Using AI response} & Explore: AI response was not inserted in the text & Report use if AI generated new ideas~\cite{acl}\\
    \cline{2-3}
    &{ AI response was inserted in the text} & Highlight text written by the AI~\cite{copyright, davis} \\
    \hline
     D3. Summary statistics &  Number of prompts used; Percentage of text written by the AI & AI-written text should not be more than 5\%~\cite{authors_guild} \\
    

     
    \bottomrule
    \end{tabular}
    \caption{\textbf{Summary of types of information required by AI-assisted writing policies.} 
    }
    \Description{A table with three rows and three columns. The first and second rows have two sub-rows for the second and third columns. The third row has no sub-rows.}
    \label{tab:typology}
\end{table*}

\section{Formative Analysis of AI-assisted Writing Policies}
\label{sec:formative}

To better understand the dimensions involved in concerns around intellectual ownership and ethical use of LLMs, we perform an analysis of what publishing outlet policies deem important to \rev{consider.} 
\rev{This included aspects such as when, how, and to what extent LLMs can be used in the creative authoring process.} \autoref{tab:typology} contains our \rev{consolidated} typology of current types of information that are important to be aware of during AI-assisted writing, \rev{per our findings}.

A tool supporting awareness and deeper understanding of these dimensions will help authors effectively and responsibly leverage powerful aids.
As a starting point, we reviewed existing guidelines for using generative models such as GPT-4 in writing from \rev{the U.S.\ Copyright Office~\cite{copyright}, The Author's Guild~\cite{authors_guild}, educational organizations such as the Modern Language Association~\cite{mla}, several creative writing publishers (e.g.,~\cite{davis}), and academic venues and publishers (e.g., ACL~\cite{acl} and ACM~\cite{acm}). We recognize that the list is not exhaustive and as we collectively learn about usage, policies will evolve. }

We conducted thematic analysis~\cite{braun2006using} to identify key dimensions from the policies.
Two authors of this paper individually reviewed the policies and open-coded the recommendations from the policies. 
\rev{The authors then met to discuss key observations and developed a codebook.
After that, the authors open-coded the policies again by following the codebook.
The inter-rater reliability between the coders was 0.91 (Jaccard's similarity).
Finally, the coders met to resolve disagreements and finalize the themes.
The full research team participated in the discussion with the coders regularly. The full list of the coded policies is available in our \textcolor{cyan}{\href{https://osf.io/gc4tr/}{OSF repository}}.}

\subsection{Patterns and Differences}

There was a consensus among policies and guidelines, \rev{irrespective of their domain (government, creative, academic, and education)}, on the need to \textbf{report} the extent of contribution from AI in the creation of content.
While specific instructions vary between the policies, all encourage authors to be transparent and disclose the use of AI.
Another shared sentiment \rev{across policies and domains} is that AI cannot be granted authorship; rather, authors should be responsible for content generated by \rev{the AI} and should acknowledge that they have themselves verified AI-generated content, and modified it where needed.

\rev{We also noticed some differences in the policies from different domains. For instance, policies from creative writing are more concerned about copyright issues and preventing LLMs from training on books without writers' permission than providing guidelines on how writers should use LLMs~\cite{authors_guild}. In comparison, policies from academic venues provide explicit guides for using and reporting LLMs~\cite{acl, acm}. Academic policies also put more emphasis on fact-checking and proper referencing~\cite{acl} than policies from creative domains. Policies from educational venues are concerned about plagiarism and academic integrity~\cite{mla}.}

Finally, there is ambiguity in the description of what needs to be reported and how writers should do that.
For example, the U.S.\ Copyright Office mentions several ways to disclose the use of AI: a brief note, acknowledgments, or providing exact contents provided by AI.
On the other hand, ACL discourages writers from using AI-generated text directly.
\textit{Nature} directs authors to report the use of AI in the method section or alternative section of the manuscript~\cite{nature_ai}, but does not explicitly specify what writers should report.



\subsection{Information Typology}

\autoref{tab:typology} presents the three major themes that emerged from the analysis---each of which corresponds to a dimension capturing the information required by AI-assisted writing policies. 
The first dimension, recurrent in the policies, separates prompts that seek new content from prompts that merely request edits to an existing text (\textbf{D1}).
Most policies recognize that we have been using computers for editorial tasks such as spelling and grammar check for a long time and writers do not need to disclose the use of LLMs for such tasks~\cite{acl}.
However, writers should report prompts that seek new content.
For example, ACM directs authors to provide a list of generative prompts used for such purposes~\cite{acm}.

All policies ubiquitously ask writers to disclose text generated by AI (\textbf{D2}).
For example, The Author's Guild specifies that\textit{``Authors shall disclose to Publisher if any AI-generated text is included in the submitted manuscript.''}
Other policies have similar clauses.
Some publishers reserve the right to reject papers that were mostly generated by AI~\cite{copyright}.
An interesting case is when an author does not directly use AI-generated text in their article, but still draws inspiration or ideas from the AI, or derives narrative angles from it.
ACL requires authors to also acknowledge such use~\cite{acl}.

While it is rare, some publishers recommend specific thresholds for using AI-written texts (\textbf{D3}).
For example, The Author's Guild restricts the authors to limit the use of AI-generated text to only 5\%~\cite{authors_guild}.
However, it does not provide any guidelines on how this 5\% should be measured in practice nor where to report these statistics. 

\section{The \toolname{} System}
\label{sec:system}

\rev{We used our formative analysis of the AI-writing policies} (\autoref{tab:typology}) \rev{to inform} the design of \toolname{}, a \rev{technology probe} that couples \rev{a text editor, an LLM, and an interaction history between the writers and LLM}.
\rev{According to Hutchinson et al.~\cite{10.1145/642611.642616}, ``Technology probes are simple, flexible, and adaptable technologies with the goal of understanding user needs and desires in a real-world setting, field-testing the technology, and motivating the design of new technologies''.
We decided on such a probe as a suitable method as it would allow us to understand how tracking provenance information can help writers, what features in \toolname{} writers find most useful, and what would be the design of future systems aiming to ensure human agency and transparency in AI-assisted writing.}
Two domain experts, who are professors of English and writing studies at our university and  co-authors of this paper, provided feedback during the development cycle of \toolname{}.
In this section, we first present \rev{the system's design rationale and then describe its details}.

\subsection{Design Rationale}

We designed \toolname{} with the following design rationales: 

\begin{itemize}
    \item[DR1]\textbf{Capture and externalize AI vs.\ human provenance.}
    The three prominent dimensions of information from AI-assisted writing policies are \textit{D1.\ prompt category}, \textit{D2.\ how writers used AI responses}, and \textit{D3.\ summary statistics}.
    Thus, our primary goal is to store this information when a writer interacts with an LLM.
    To support provenance, the system should externalize this information to the writer in an easily understandable format.
    The writer should then be able to \rev{freely go back and forth} in the interaction history. 

    \item[DR2]\textbf{Integrate provenance in artifact.}
    Externalizing interaction history may not be enough.
    To make sense of the history, the system should connect the artifact (i.e., the text document) with the history~\cite{DBLP:conf/ACMdis/HoqueGE22, DBLP:conf/ACMdis/HoqueGE23}.
    Moreover, some policies require authors to highlight text written by AI in the artifact itself.
    Thus, content generated by \rev{the AI} should be clearly visible in the text and linked to the information stored for DR1.

    \item[DR3]\textbf{Integrate writer's judgment in provenance.}
    While we aim to store user interaction automatically, we may not be able to store all information automatically. 
    For example, one category in \autoref{tab:typology} references when a writer takes inspiration from AI outputs, but does not use the output directly. 
    Since the influence here is implicit and difficult to quantify, we decided to facilitate mechanisms for writers to integrate this information in the interaction history.
    To improve agency for writers, we also decided that writers should be able \rev{to edit or modify the interaction history if needed}.
    One implication of this decision is that our tool is not a tool to enforce AI-writing policies; rather it is a tool for writers to be able to measure their own compliance, while being able to design disclosures and be transparent.

    \item[DR4]\textbf{Extensible/flexible.}
    While existing policies provided a baseline for our work, they are still evolving.
    The technology around LLMs is also particularly fluid.
    Thus, our design should be extensible to new requirements, if they appear in the future.

\end{itemize}

\subsection{Visual Interface}

\autoref{fig:teaser} shows the full interface for \toolname{}.
Because of the rich dependencies between text, prompts, and interaction history, we opted for a visual approach with data visualization components.
The interface is divided into three modules: a) a rich text editor; b) an interface to interact with GPT-4; and c) a module to visualize interaction history with LLMs.
\rev{By default, we allocate 1/3 of the screen width to each module.
However, we provide toggle buttons to hide or show each module.
Whenever a writer toggles the visibility of a module, we redistribute the available width to the visible modules equally.
For example, a writer can hide the GPT-4 and visualization module to write ``distraction-free''.}
We describe individual components of \toolname{} below, and include a video of the tool as supplementary material.

\begin{figure*}[tbh]
    \centering
    \includegraphics[width=0.9\textwidth]{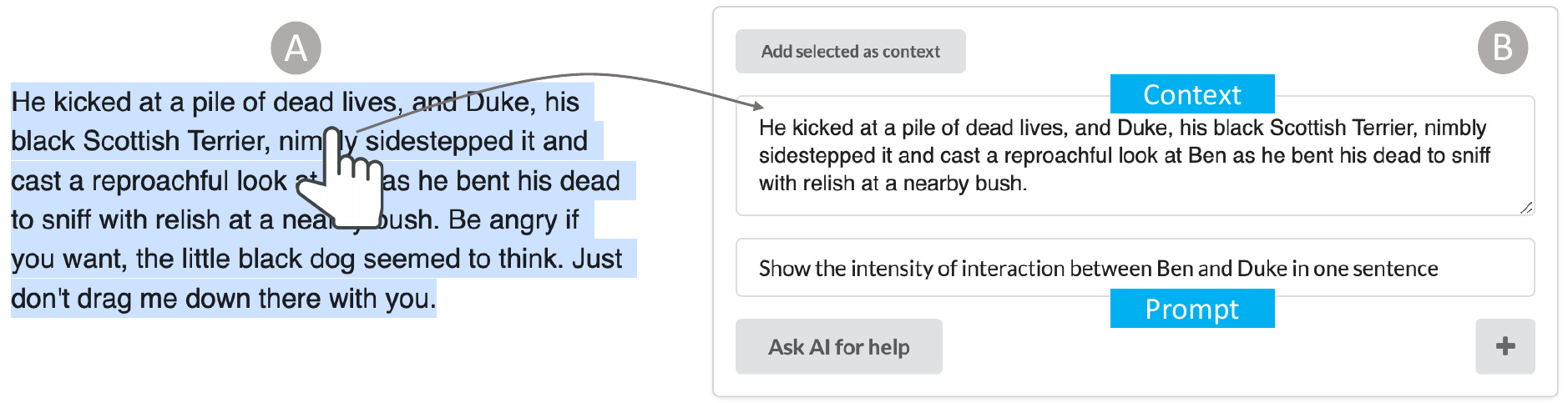}
    \caption{\textbf{Prompting GPT-4 in \toolname{}.}
    A) By highlighting any portion of the text in the text editor, the user can select that text as context for prompting GPT-4.
    B) The selected text is automatically pasted into the context box.
    The user can specify the task to perform in the prompt box.}
    \label{fig:prompting}
    \Description{The figure has two panels. The left panel shows a paragraph of text content, and the right panel shows several input boxes and buttons for further processing of the input.}
\end{figure*}

\begin{figure*}[tbh]
    \centering
    \includegraphics[width=0.9\textwidth]{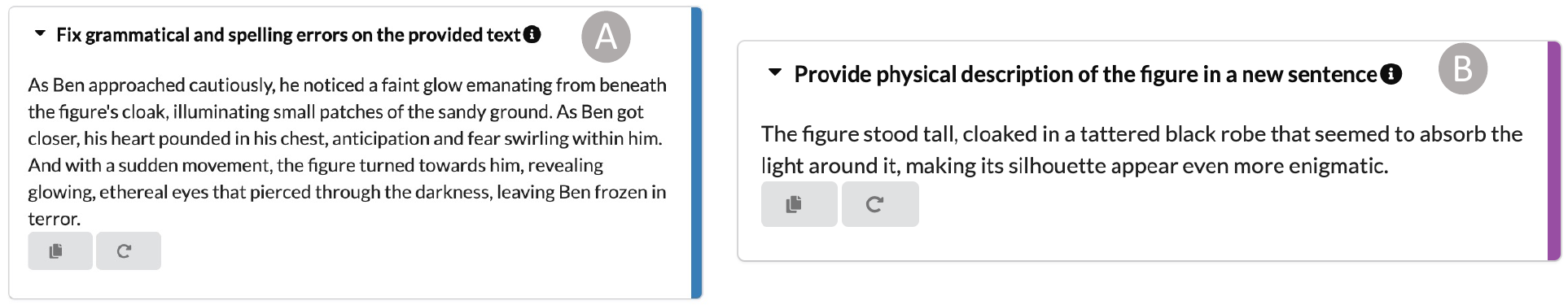}
    \caption{\textbf{Design of the prompt card.}
    We encapsulate each prompt and AI response in a card.
    The title shows the prompt.
    Users can hover over the information icon to see the context.
    Each card contains a \faCopy~copy button and \faRepeat~redo button for regenerating the AI response.
    We categorize each prompt as either seeking new contents (blue) or seeking editorial help (purple) on an existing text. For instance, 
    A) shows a prompt seeking new content, and B) is a prompt seeking editorial help.}
    \label{fig:prompt_card}
    \Description{The figure has two panels. Both panels show paragraphs of text content. The left panel shows a paragraph of text content after fixing grammatical errors, and the right panel shows a paragraph describing a figure.}
\end{figure*}

\subsubsection{Prompting LLMs}

The prompting interface in \toolname{} bears similarities with the current ChatGPT interface.
It has a text box for writing the prompt and an optional text box for specifying the context of the prompt (\autoref{fig:prompting}B).
A user can highlight a portion of the text in the editor to be automatically selected as an additional context for prompting the AI (\autoref{fig:prompting}).
The response from an LLM such as GPT-4 gets appended below the text boxes.
The prompt wizard suggests several standardized creative composition interactions, such as ``summarize,'' ``elaborate,'' ``enumerate,'' ``introduce,'' and ``conclude.'' 
A writer can write a free-form prompt or choose one from the standardized recommendation. 

\subsubsection{Prompt Card}

We encapsulate each prompt and the relevant AI response in a card, the popular UI component for designing modular objects (Figure~\ref{fig:prompt_card}).
As per \textbf{DR1}, we categorize each prompt as either seeking 
\begin{imageonly}\promptgenerate{generation of new contents}\end{imageonly}
 or
  \begin{imageonly}\promptedit{editorial help on an existing text}\end{imageonly}
  .
We use the following soft prompt with the actual prompt to identify the category: 
\texttt{``For the input text, reply `Edit' or `Generate' if the text intends to edit existing text or generate new text.
Consider paraphrasing an existing text, or grammatical and spelling check as an Edit. 
Input sentence - '' + input prompt.}
\rev{To validate the performance of this method, we created a dataset of 150 prompts.
Two authors of this paper collaboratively created the prompts and then labeled them as either targeted at 
\begin{imageonly}\promptgenerate{generating}\end{imageonly}
 new content or focused on 
 \begin{imageonly}\promptedit{editorial}\end{imageonly}
  support of existing content. 
We ensured that the prompts spanned a wide range of writing compositions and were challenging to decode.
We then measured the accuracy of the soft prompt in classifying the prompts correctly.
The accuracy was 96\%. The list of prompts and their labels are available in the 
\textcolor{cyan}{\href{https://osf.io/gc4tr/}{OSF repository}}. }
We encode the prompt category at the right border of the card body with either \texttt{purple} color for indicating generation or \texttt{blue} for indicating edit.

\begin{figure*}[t]
    \centering
    \includegraphics[width=0.9\textwidth]{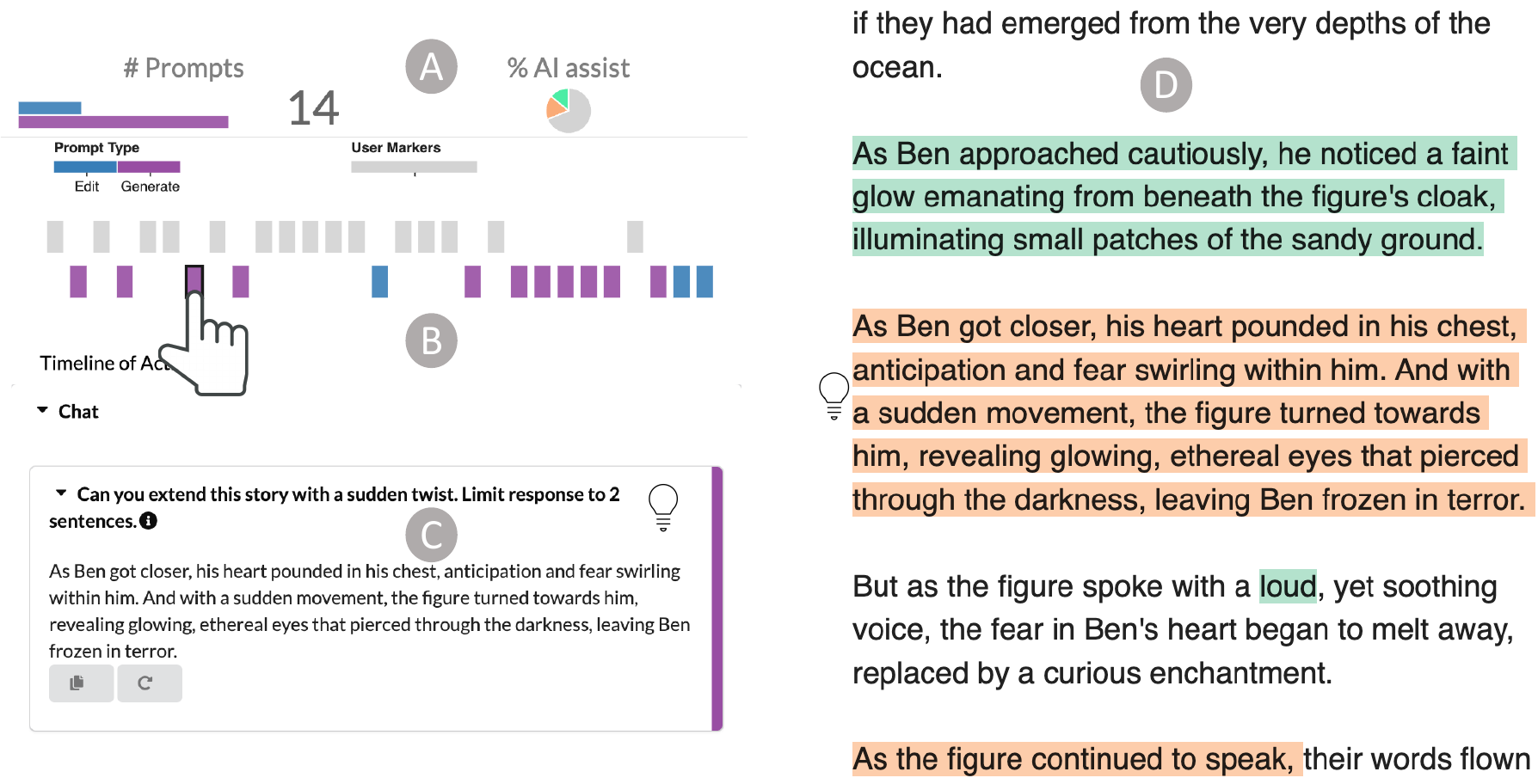}
    \caption{Visualization and interaction in \toolname{}.
    A) Summary statistics: number of prompts and percentage of assistance from AI.
    B) The timeline shows the prompts (blue or purple tiles) in the context of the user's writing behavior (e.g., writing a new sentence).
    Hovering over a colored tile will show the respective (C) prompt and text highlighted in the text editor (D).}
    \label{fig:ui}
    \Description{The figure has two panels. The left panel contains some visual elements, and the right panel shows a text editor.}
\end{figure*}

\begin{figure*}[t]
    \centering
    \includegraphics[width=0.9\textwidth]{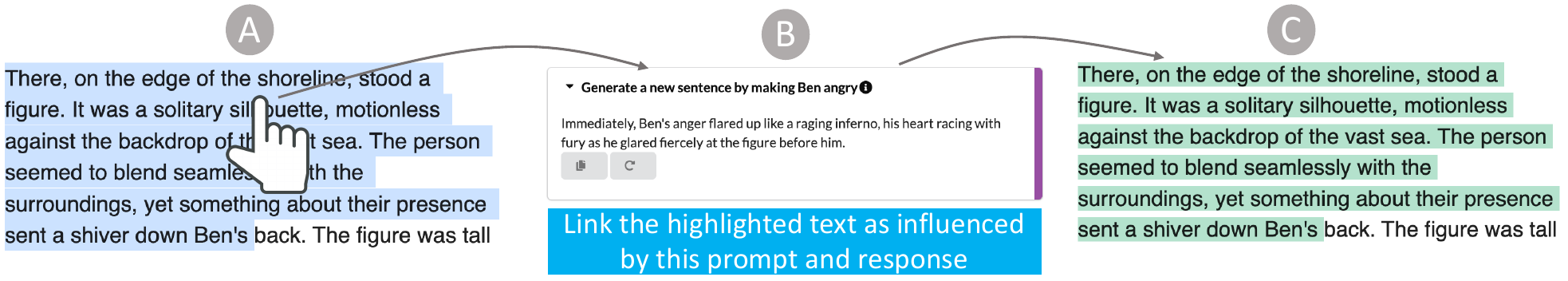}
    \caption{\textbf{Manually linking a portion of the text with a prompt in \toolname{}.}
    A) The user highlights a portion of the text.
    B) The user can link the text with a prompt from the prompt history.
    The user can either label it as 
    \begin{imageonly}\aiwritten{AI-written}\end{imageonly}
     or 
     \begin{imageonly}\aiinfluenced{AI-influenced}\end{imageonly}
     .
    In this case, the writer labels it as 
    \begin{imageonly}\aiinfluenced{AI-influenced}\end{imageonly}
    .
    C) The text color changes to green to indicate the change in the label.}
    \label{fig:manual_linking}
    \Description{The figure has three panels. The middle panel contains a paragraph of text content, and both the left and right panels show a text editor.}
\end{figure*}

\subsubsection{Visualizing AI vs.\ human provenance}
\label{sec:viz_prov}

We externalize the provenance information using several interactive visualizations (\textbf{DR1}). We visualize two summary statistics in \toolname{} (Figure~\ref{fig:ui}A)\rev{: the counts for the prompts in a bar chart; and in a pie chart, the percentages for 
\begin{imageonly}\aiwritten{AI-written}\end{imageonly}, 
\begin{imageonly}\aiinfluenced{AI-influenced}\end{imageonly}, and text written by the writer.}
\rev{We consider text written by the AI and then used verbatim to be as \begin{imageonly}\aiwritten{AI-written}\end{imageonly} 
and highlight them with an \texttt{orange} color in the text editor and pie chart.
When a user copies full or parts of the response from the prompt card and pastes it into the text editor, we automatically highlight that text as \begin{imageonly}\aiwritten{AI-written}\end{imageonly}
, update the pie chart, and link the prompt with the text (\textbf{DR2}).
If a user re-writes or edits a portion of the 
\begin{imageonly}\aiwritten{AI-written}\end{imageonly}
 text, we remove the highlight from the specific portion of the text and mark it as written by the user. The other parts of the text remain as 
 \begin{imageonly}\aiwritten{AI-written}\end{imageonly}.}

\rev{Writers can manually mark any text as 
\begin{imageonly}\aiinfluenced{AI-influenced}\end{imageonly}
 (\textbf{DR3}). For instance, after re-writing an \begin{imageonly}\aiwritten{AI-written}\end{imageonly}
 text, writers who feel that the text is influenced by the AI can mark the text as such. Following \textbf{DR3}, we do not automatically detect 
 \begin{imageonly}\aiinfluenced{AI-influenced}\end{imageonly}
  text, and rather leave it to the writers' discretion and judgment. This said, flagging potential \begin{imageonly}\aiinfluenced{AI-influenced}\end{imageonly}
   text for writers to critically review could still be useful. To this end, we explored several automatic methods aimed at identifying 
   \begin{imageonly}\aiinfluenced{AI-influenced}\end{imageonly}
    text.
For instance, following Kim et al.~\cite{Kim2023cells}, we experimented with the BLEU score---a measure used to evaluate the similarity between a piece of text and references.
We calculated the BLEU score between text re-written by a person with the responses from LLMs to see if we can reliably detect 
\begin{imageonly}\aiinfluenced{AI-influenced}\end{imageonly}
 text.
We also experimented with OpenAI's classifier for detecting AI-written text~\cite{openai_detector}.
We decided against using these methods as our domain experts did not find the methods to be consistent and sufficiently reliable. OpenAI also lists the shortcomings of such methods~\cite{openai_detector}. We determined that
wrong predictions and interpretations could negatively impact the user's experience, trust, and agency in the user study, and decided to omit the feature in the current probe. }


The timeline shows interaction history in a linear fashion (Figure~\ref{fig:ui}B), using a colored rectangle (either blue or purple) for each prompt.
We insert a new grey rectangle in the timeline each time the user writes a new line to show writing activity in comparison to prompting AI.
The timeline is scalable to more variables and information as we can encode the new information in a new row in the timeline (\textbf{DR4}).
The timeline also extends horizontally when the rectangle width becomes less than a threshold (default 5px) and provides a scroll bar to see the extended content.

A user can hover over the colored rectangles to see the linked prompts (Figure~\ref{fig:ui}) and the linked text, if any (\textbf{DR2}). By clicking any colored rectangle, the user can keep the linked prompt and text highlights stay visible. 

\subsubsection{Linking Visualization and Artifact}

We used QuillJS~\cite{QuillJS} as a rich text editor \rev{in our interface, where a user can read or write textual content, and apply traditional formatting. Beyond these traditional operations, a user can also perform the following actions related to AI-assisted writing in the text editor}:

\begin{itemize}
    \item\textbf{Manually label text.}
    Following \textbf{DR3}, a user can select a portion of the text and then label the text as either \begin{imageonly}\aiwritten{AI-written}\end{imageonly} or 
    \begin{imageonly}\aiinfluenced{AI-influenced}\end{imageonly}
     using a button named \texttt{Highlight}.
    Additionally, the user can link a prompt from prompt history with the highlighted text (\autoref{fig:manual_linking}).
    This helps writers to manually annotate any text in the case where our system cannot automatically annotate them. 

    \item\textbf{Manually remove label.}
    In a similar manner, writers can remove \rev{annotations (
    \begin{imageonly}\aiwritten{AI-written}\end{imageonly} or 
    \begin{imageonly}\aiinfluenced{AI-influenced}\end{imageonly}
    )} and links with prompts by first highlighting a portion of the text and then clicking a button named \texttt{Unhighlight}.

    \item\textbf{Click on annotated text.}
    A user can see the linked prompt to an annotated text by clicking on it (\textbf{DR2}).

\end{itemize}

\section{Evaluation}
\label{sec:eval}

We conducted a user study with 13 creative writers.
The overarching goal of the study was to \rev{answer our original high-level research question}:
\textit{How \rev{can} externalizing provenance information help AI-assisted co-writing?}
We used \toolname{} as a \rev{technology probe} to explore this question.
Since provenance can impact many facets of AI-assisted writing (Sections \ref{sec:intro} and \ref{sec:related-work}), we seek to answer the following specific research questions (RQs):

\begin{itemize}
    \item[\textbf{RQ1:}] How does \toolname{} affect \rev{a writer's} interaction with \rev{an LLM}?
    \item[\textbf{RQ2:}] How does \toolname{} affect \rev{a writer's} ownership concerns while \rev{receiving AI-writing support} from \rev{an LLM}?
    \item[\textbf{RQ3:}] How does \toolname{} help \rev{a writer to} communicate \rev{the extent of} their use of LLMs?
    \item[\textbf{RQ4:}] How does \toolname{} help \rev{a writer to} conform to policies on AI-assisted writing?
\end{itemize}

\subsection{Different Forms of Writing}

While \toolname{} is generalizable to different types of writing (e.g., creative, argumentative, academic), each type has different goals, tasks, and styles.
Instead of recruiting writers from diverse domains, we decided to conduct a case study with creative writers. \rev{We thus note that while our evaluation might inform the adoption of \toolname{} in other domains, our results remain specific to our chosen use case.} 

Creative writers \rev{commonly use a wide range of techniques that are regarded as integral to creative, expressive writing}.
\rev{These include vivid, concrete language; metaphors and similes; syntactical variety; alliteration; and other literary devices.}
Furthermore, the impact of LLM \textit{hallucination}~\cite{ji2023survey} is less problematic for creative writing than non-fiction and academic writing, thus removing a potential confound in the study.
In addition, there has been a significant backlash from creative writers about the use of AI in creative writing, instrumented by the recent strike from screenplay writers and artists in Hollywood.
LLMs directly threaten their bread and butter.
Previous studies also reported that creative writers have ownership and agency issues when using LLMs~\cite{DBLP:conf/chi/MirowskiMPE23, yuan2022wordcraft}.
Given the premise of this work, creative writers \rev{are perfectly suited} to help us answer the \rev{research questions}.

\subsection{Study Conditions}

We conducted a repeated-measures within-subject experiment with the following two conditions (counterbalanced):

\begin{itemize}
    \item[\textbf{C1.}] \textsc{Baseline}: A ChatGPT-like interface with a text editor.
    Participants are able to write, use GPT-4, and see a list of prompts and responses from GPT-4 in a sidebar.
    We include a screenshot of the baseline in the supplement. 
    \item[\textbf{C2.}] \textsc{\toolname{}}: Our tool with all interactive support.
\end{itemize}

We decided that a text-only interface is best suited here as a baseline as it will represent the current LLM-based writing interfaces (e.g., ChatGPT and Google Docs). It will help us capture the impact of provenance visualization in \toolname{}.

\subsection{Participants}

We recruited participants by advertising in our university's Writing Center \rev{as well as} English, Literature departments.
\rev{Our participants varied in terms of self-reported gender (male = 5, female = 7, prefer not to say = 1, other = 0), age (min = 19 years, max = 56 years, mean= 26 years, SD = 4.2 years), experience in writing different creative materials (fiction, non-fiction, short stories, and poems), and years of experience as creative writers (min = 5 years, max = 26 years, mean = 9.2 years, SD = 4.3 years).}
All participants had published works in their portfolio.
Participants received a \$40 USD gift card for their time.

\rev{All participants reported} prior experience in using LLMs (e.g., ChatGPT) or were aware of their \rev{use} in creative writing, \rev{but none mentioned LLMs to be a critical part of their writing process}.
\rev{Five participants reported using ChatGPT infrequently for various editorial writing tasks such as rephrasing a text or changing the mood of the text.
Two participants had used ChatGPT to explore different narrative angles.
Other participants had tested ChatGPT out of intellectual curiosity.
Two participants had actively participated in the 2023 WGA/SAG-AFTRA strike.}

\subsection{Tasks}

It is difficult to design tasks with objective goals for creative writers~\cite{DBLP:conf/ACMdis/HoqueGE22, DBLP:conf/ACMdis/HoqueGE23}.
Their work typically does not adhere to predefined structures and depends on their artistic styles and idiosyncrasies~\cite{DBLP:conf/ACMdis/HoqueGE23}.
Thus, we decided to ask writers to write short stories using our interfaces for a fixed amount of time (20 mins) while being able to prompt GPT-4.
We aimed to study their interaction with the two study conditions and collect feedback through semi-structured interviews to answer the RQs.

\subsection{Measures}

Since the study tasks did not involve any objective goals, we opted for a qualitative methodology.
Another reason for this choice is that concepts relevant to our study (e.g., agency, transparency, ownership) are mostly abstract concepts and are difficult to operationalize quantitatively~\cite{yuan2022wordcraft}. 
Instead, we designed a semi-structured interview for capturing writers' feedback.
We asked writers about how the study conditions impacted their interaction with LLMs, agency, control, and ownership.
\rev{We also asked writers about the usefulness of each interface to support communication and transparency around AI-assisted writing.}
The interview script is available in the \textcolor{cyan}{\href{https://osf.io/gc4tr/}{OSF repository}}.

We also asked participants to rate the study conditions on a 7-point Likert scale across three subjective dimensions:

\begin{itemize}

    \item \textit{Ownership}: On a scale of 1 (not at all comfortable) to 7 (very comfortable), how comfortable would you be in publishing the short story under your name?
    
    \item \textit{Communication and Transparency}: On a scale of 1 (not helpful at all) to 7 (very helpful), \rev{how helpful would the tool be to you} in communicating your use of AI to others (e.g., publishers, readers) for transparency?
    
    \item \textit{Conformity}: \rev{On a scale of 1 (not helpful at all) to 7 (very helpful), how helpful would the tool be to you in conforming to the given AI-assisted writing guideline?}
    
\end{itemize}

Finally, following prior literature~\cite{DBLP:journals/corr/abs-2212-09746, Kim2023cells}, we asked participants to rate each condition on a 7-point Likert scale (1: strongly disagree, 7: strongly agree) across the following six dimensions for capturing the usability of LLM support:

\begin{itemize}
    \item \textit{Helpful}: ``I found the AI helpful.''
    \item \textit{Ease}: ``I found it easy to write the advertisement.''
    \item \textit{Experiment}: ``I felt that I experimented with various ideas and generated alternatives.''
    \item \textit{Iteration}: ``I felt that I iterated various times on ideas and the generation process.''
    \item \textit{Pride}: ``I am proud of the final output.''
    \item \textit{Unique}: ``The story I wrote feels unique.''
\end{itemize}

\subsection{Procedure}

Before each session, we asked participants to familiarize themselves with the policy on AI-assisted writing from the U.S.\ Copyright Office~\cite{copyright}.
We also asked participants to think about the plots and settings for two short stories, but asked them not to start writing in advance of the research study session.
Each session started with participants signing the consent form and a brief introduction about the goal of the study from the study administrator.
After that, we introduced the first study condition (tool)  with a brief demo.
We encouraged participants \rev{to ask questions at this stage and then to explore} different features of the tool \rev{using} a training story.

Participants then started the first writing session (20 minutes) \rev{in which they were asked to write a short story}.
We clarified to the participants that they \rev{did not need to finish the full story}; rather this is a timed experience. 

At the end of the \rev{first} writing session, participants filled out a survey to provide their subjective experience.
Participants \rev{then started the second writing session where they were asked to write their second story using the other interface}, following the same procedure as for the first condition.
\rev{We followed this with a second survey and then concluded the study} with a semi-structured interview.
\rev{During the interviews, participants shared their experience with both the baseline and the tool conditions and discussed their use of the LLM in the writing process.}

\subsection{Analysis Plan}

Similar to our formative analysis (Section~\ref{sec:formative}), two authors of this paper independently open-coded the anonymized post-study interview transcripts and then conducted a thematic analysis.
The coders met regularly to discuss and refine the codes and themes.
The coders also discussed the codes and themes with the entire research team.
\rev{The initial inter-rater agreement was 0.86 (Jaccard's similarity).}

For quantitative measures and subjective ratings, we decided to avoid traditional null-hypothesis-based statistical testing in favor of estimation methods to derive 95\% confidence intervals (CIs) for all measures~\cite{dragicevic2016fair}.
We employed non-parametric bootstrapping with $R = 1,000$ iterations.
We also report the standardized effect size (Cohen's $d$).

\begin{figure*}[tbh]
    \centering
    \includegraphics[width=0.95\textwidth]{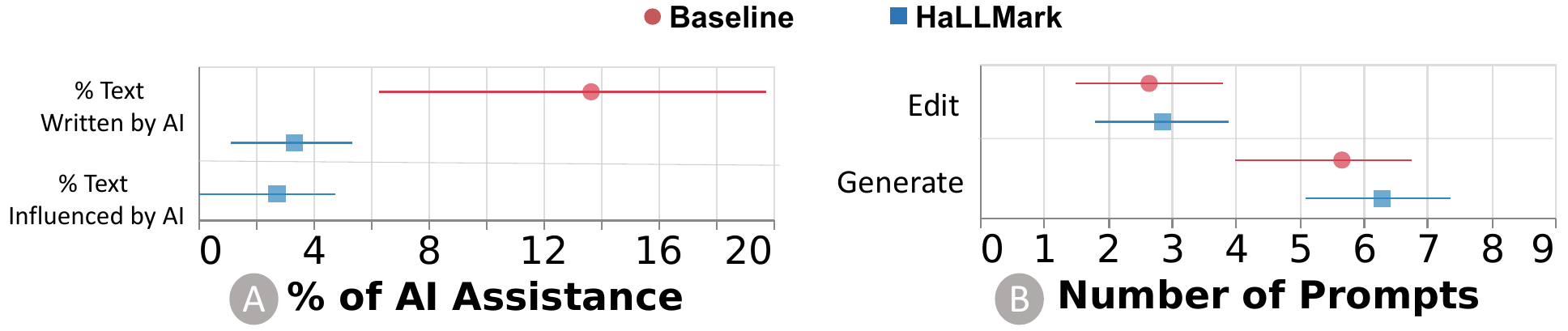}
    \caption{\textbf{Percentage of AI assistance and number of prompts and while using the baseline tool and \toolname{}.}
    Error bars show 95\% confidence intervals (CIs).
    The baseline condition did not have the option to label text as 
    \begin{imageonly}\aiinfluenced{AI-influenced}\end{imageonly}
    .
    Thus, we see only one mark for that category in Figure A.}
    \Description{A Figure with two sub-figures. Each figure shows some colored lines, rectangles, and circles. }
    \label{fig:ai_contributions}
\end{figure*}

\subsection{Results}
\label{sec:results}


\subsubsection{RQ1: Interaction with LLM}

We found that \toolname{} significantly changed the writers' interaction with the LLM\rev{ compared to the baseline}.
In the post-study interviews, participants mentioned that \toolname{} instilled a sense of awareness and encouraged them to actively evaluate AI-assistance from the beginning of the process (P1-4, P7, P10-13).
For example, P2 and P8 said,

\begin{quote}
    \textit{``I liked [\toolname{}] better because I was trying to use the AI without overusing it, and there were times when I felt like I was [doing that].
    But then it said, `Oh, you know, 90 or 95\% of this writing is yours,' so, you know, more than I thought.
    So that was nice to have, and I liked having that information all the time.''}
    (P2)
\end{quote}

\begin{quote}
    \textit{``I was keeping an eye on the text highlighted by yellow color and the percentage of that in the pie chart.
    It certainly made me conscious and encouraged me to modify text generated by the machine.''} (P8)
\end{quote}

On the flip side, some participants mentioned that it is possible that the tool may make some writers nervous and overly conscious about overusing the LLM, particularly in the eyes of readers and publishers, or even other writers (P5-6).
This stigma can hamper their creative process. 

We found evidence of the impact of \toolname{} in the percentage of \begin{imageonly}\aiwritten{AI-written}\end{imageonly}
 text in the final stories. \rev{As defined in section~\ref{sec:viz_prov}, we consider text directly generated by GPT-4 as 
 \begin{imageonly}\aiwritten{AI-written}\end{imageonly} and exclude text that was generated by GPT-4 but later re-written by the writers.
We also measured text marked as \begin{imageonly}\aiinfluenced{AI-influenced}\end{imageonly} by the writers.} 
\autoref{fig:ai_contributions}A shows the percentage of text written and influenced by the AI in the final document for the two conditions.
On average, the stories contained 13.66\% (CI = [6.30, 19.76]) text written by the AI when participants used the baseline.
In comparison, the stories contained only \rev{3.48\% (CI = [1.23, 5.26])} text written by AI when participants used \toolname{}. 
Additionally, participants labeled 2.99\% (CI = [0.00, 4.77]) of the total text as \begin{imageonly}\aiinfluenced{AI-influenced}\end{imageonly} in the stories written using \toolname{}.

However, we did not observe any difference in the number of prompts used in the two conditions.
Regardless of the condition, participants preferred asking the AI to generate new content.
\autoref{fig:ai_contributions}B shows the number of prompts used by participants in the two conditions.
On average, participants used 2.65 (CI = [1.50, 3.80]) prompts seeking editorial help with the baseline.
Participants used a similar amount of editorial prompts (2.86 with CI = [1.80, 3.89]) while using \toolname{}.
The small effect size of 0.04 (Cohen's $d$) indicates no practical difference \rev{between the conditions}. 

Participants used prompts seeking generation more frequently than editorial prompts.
On average, participants asked 5.66 (CI = [4.00, 6.76]) prompts seeking new content while using the baseline.
Participants used a similar amount of prompts seeking new contents (6.29 with CI = [5.10, 7.37]) while using \toolname{}.
The small effect size of 0.1 (Cohen's $d$) indicates a very small practical difference \rev{between the two conditions}.


\begin{figure*}
    \centering
    \includegraphics[width=0.95\textwidth]{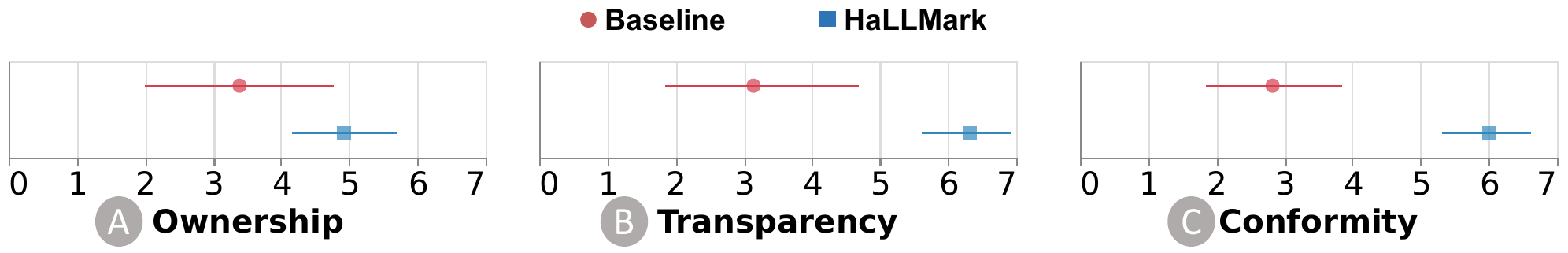}
    \caption{\textbf{Self-reported subjective ratings for ownership, transparency, and conformity of AI policies.}}
    \label{fig:con_trans_own}
    \Description{A Figure with three sub-figures. Each figure shows some colored lines, rectangles, and circles. }
\end{figure*}

\subsubsection{RQ2: Agency and Ownership}

We found evidence that the situational awareness provided by \toolname{} improved writers' control over the process.
As a result, writers were able to measure their contribution better when using \toolname{}.
P7's comment below summarizes their experience,

\begin{quote}
    \textit{``[\toolname{}] made me feel less confused in a way, even to myself, like what did I generate?
    What did the AI generate?
    What was influenced by the AI?
    It felt easy to apply the green highlighting for what was influenced, and the fact that it just automatically applying the orange highlighting for what the AI had generated felt pretty seamless.
    And it gave me sort of this feeling of reassurance and control that I did not find in the [baseline] interface.''} (P7)
\end{quote}

Using \toolname{}, some participants were able to perceive AI as a collaborator, rather than as an external agent (P3, P9, P13). For example, P13 said, 

\begin{quote}
    \textit{``With all the information showing my work and AI's work, it felt less robotic and more like I was collaborating with someone.''} (P3)
\end{quote}

Of course, there is danger inherent with anthropomorphizing AI~\cite{Deshpande2023, Shneiderman2022}; AI models are not persons and thus cannot be authors in the true sense, and there are legal, safety, security, trust, and reliability concerns in such relationships~\cite{Epstein_2023, Shneiderman2022}.

The overall positive experience was reflected in the  \textit{ownership} ratings provided by the participants (\autoref{fig:con_trans_own}A).
On average, the rating for the baseline was 3.39 (CI = [2.00, 4.76]), and for \toolname{}, the rating was 4.92 (CI = [4.15, 5.69]).
A standardized effect size of 0.46 (Cohen's $d$) indicates a medium effect of the study condition.

\subsubsection{RQ3: Communication and Transparency}

Most participants preferred \toolname{} to communicate the extent of AI contributions in the final artifact.
According to P1,

\begin{quote}
    \textit{``I do not even see how I can use the first one [baseline] for communicating.
    [With \toolname{}], you can literally copy the text with colors and send it to someone in seconds.
    You can send the pie chart and the rectangles for more breakdowns.''} (P1)
\end{quote}

However, two participants had reservations \rev{against} using \toolname{} for communication.
P8 was worried that people might ``nitpick'' their writing if it was completely transparent and that readers would harshly criticize the use of AI.
P4 preferred the timeline and text highlighting for communicating AI contributions, but did not want to share the summary statistics.
They felt that readers might reduce their work to a single number (e.g., only 80\% of the text was written by the author).
The timeline would presumably show their contribution more clearly. 

\autoref{fig:con_trans_own}B shows the participants ratings for how useful \toolname{} they felt it is for communication and transparency.
On average, the rating for the baseline was 3.14 (CI = [1.85, 4.69]).
The average rating for \toolname{} was 6.31 with CI = [5.61, 6.92].
The standardized effect size ($d$ = 2.27) shows a very large effect of the study condition.

\subsubsection{RQ4: Conformity to AI-assisted Writing}

All participants preferred \toolname{} for evaluating conformity to AI-assisted writing policies.
For example, P5 said, 

\begin{quote}
    \textit{``When I read the policy before the study, I was like `Ooh! this will be such a pain in the [posterior\footnote{\textit{Equus asinus.}}].'
    But, then when I used the tool, I was like, `Okay, this is easy!'
    I would totally use the second tool [\toolname{}] if I needed to follow a policy like this.''} (P5)
\end{quote}

We noticed a large difference in the subjective rating for this dimension for the two conditions (\autoref{fig:con_trans_own}C).
On average, the subjective rating for the baseline was 2.83 (CI = [1.85, 3.85]).
The average rating for \toolname{} was 6.00 with CI = [5.31, 6.61].
The standardized effect size ($d$ = 2.10) indicates a very large effect of the study condition.

\begin{figure*}[tbh]
    \centering
    \includegraphics[width=0.90\textwidth]{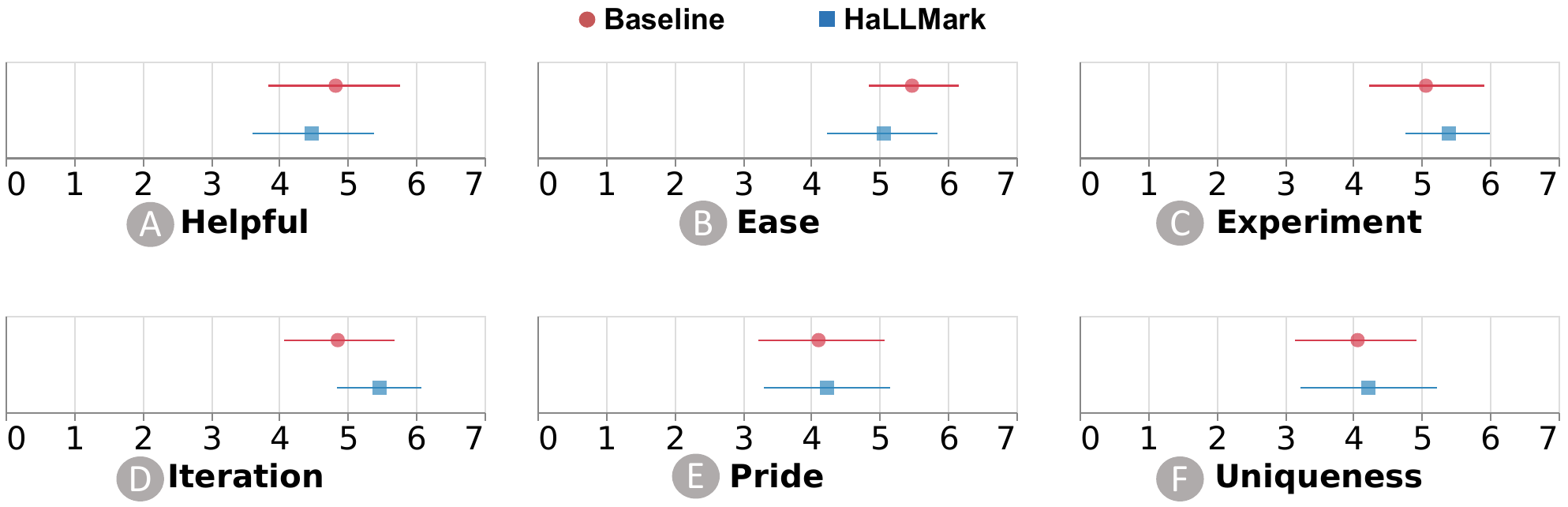}
    \caption{\textbf{Subjective perception of LLM support.}
    We collected subjective perception for LLM support across six dimensions~\cite{DBLP:journals/corr/abs-2212-09746}.}
    \label{fig:usability_metrics}
    \Description{A Figure with six sub-figures. Each figure shows some colored lines, rectangles, and circles. }
\end{figure*}

\subsubsection{Subjective Perception of LLM Support}

\autoref{fig:usability_metrics} shows participants' subjective ratings on the usability of LLM support~\cite{DBLP:journals/corr/abs-2212-09746, Kim2023cells}.
Similar to previous research~\cite{Kim2023cells}, we did not observe any significant difference in these dimensions.

\subsubsection{Cognitive Load and Usability}

Overall, participants found \toolname{} to be easy to use.
\rev{Prompting LLMs is a relatively new activity for writers.
\toolname{} added an extra task on top of prompting: tracking and verifying provenance information.
However, participants did not report any excessive cognitive load due to this task in the post-study interviews.
We believe there are three reasons for that.
First, the general feedback from all participants indicates that verifying provenance information is a real need for writers who want to use LLMs and writers do not see this as an extra task.
Second, several participants appreciated the use of visualization to seamlessly integrate the tracking task in \toolname{}.
Participants used words such as ``easy-to-understand'', ``simple'', and ``cool'' to describe the visualizations and interaction. }
\rev{Finally, participants found the ``distraction-free'' mode useful to focus on specific tasks.
We noticed several participants turned on and off the three modules of \toolname{} in different combinations to switch between writing, prompting, and validating provenance information.
For example, P6 turned off the GPT-4 and visualization modules whenever they were writing, turned on the GPT-4 module for prompting the LLM, and  then turned on the visualization module for seeing the summary statistics and prompt history. }

\rev{Participants also suggested several improvements to \toolname{}.
P3 and P6 suggested adding an option to turn on and off the text highlighting, as they can become distracting to writers for long-term use.
In the current implementation, the text highlighting is always on. }
P9 wondered if they could add notes to the prompt or the text editor directly.
This might be useful for providing an explanation if needed.
P1 asked for more control over prompt generation, as for instance controlling the randomness of the text generation.


\section{Discussion, Limitations, and Future Work}
\label{sec:disc}

Our \rev{findings show} that capturing and externalizing provenance information have a significant impact on how writers interact with LLMs \rev{as well as on} their agency and control over the process.
\rev{Our writer participants used \toolname{} to easily track the AI's contribution} with respect to their own contribution.
The tool helped writers maintain a level of AI contribution that they were comfortable with.
This provided a sense of control in \rev{the writers' minds} and improved their ownership of the final artifact.
Feedback from \rev{our participants} indicate that \toolname{} \rev{can help them become more transparent} about the co-writing process and conform to AI-assisted writing policy without manually preparing disclosures. 
Below we discuss the broader impacts of our work on interactive and intelligent writing support tools.

\subsection{Normalize Transparency and Accountability in AI-assisted Writing}

Interactive provenance information helped writers make informed decisions while producing their short stories.
While writers were generally enthusiastic about sharing the provenance information with readers, some writers were not comfortable reporting the AI-generated parts publicly.
They thought that disclosing the use of LLMs might make them susceptible to criticism.
The fear of criticism is understandable, as \rev{it might diminish the perceived contribution of the human writer} and presumably lead to disapproval of their creativity~\cite{stokel2022ai}.
We note that such concerns are not unique to the use of intelligent tools---similar taboos exist for the use of reference materials for inspiration~\cite{holinaty2021supporting}.

We believe the right way to remove the stigma around AI-assisted writing is by encouraging writers to be more transparent and accountable, and democratize tools to support this goal~\cite{liao2023ai, hosseini2023ethics}.
However, this also requires that readers, publishers, and other writers \rev{become charitable and open-minded} about LLMs going forward.

\subsection{Design Implications}
\label{sec:implications}

\subsubsection{Writers Want to Use LLMs for Content Generation, not Editing}

We found that participants in the study were mostly interested in generating new content or ideas (not the whole story) using GPT-4.
Although existing policies indirectly encourage writers to use the LLM for editorial purposes~\cite{acl}, writers did not find that useful during the study.
Rather, they were intrigued by its generation power and wanted to use it for overcoming challenges such as writer's block, difficulties in expressing nuanced and expressive details about a new scene, or taking narrative inspiration.
\rev{While this result is in line with several previous studies~\cite{DBLP:conf/chi/MirowskiMPE23, DBLP:conf/ACMdis/GeroLC22}, it contrasts with a survey that found that 60\% of the surveyed writers want to use LLMs for editorial purposes~\cite{ai_survey}.}

\rev{One explanation behind this contrasting result could be that we collect writers' feedback based on the experience with tangible interfaces whereas the survey depended on writers' preconceived perceptions of LLMs.
It is also possible that writers' perceptions have changed since the time of the survey (May 2023) due to the introduction of policies on AI-assisted writing.
Another caveat here is that the writing sessions in our study were short (20 mins) and likely do not fully capture how writers might use tools such as \toolname{} for long-form writing (e.g., fiction).
Nevertheless, we believe that, in the future, organizations and authorities will likely benefit by focusing on devising policies for ethical and responsible content generation, rather than limiting writers to the use of LLMs for editorial purposes.}

\subsubsection{When to Write, When to Prompt, and When to Verify?}

\rev{Our findings indicate that when and how writers want to switch between writing, prompting, and tracking provenance depends on writers' personal style, needs, and idiosyncrasies.
During the study, we noticed several participants turned on and off the three modules in different combinations.
This observation indicates that future AI-assisted writing tools will benefit from allowing writers to freely switch between writing and AI-related tasks.
In the future, we aim to conduct a longitudinal study to comprehensively understand how the tracking task impacts writing and when and how writers want to verify provenance information. Recent studies on understanding writers' needs for AI support are inspiring in this scenario~\cite{DBLP:conf/chi/GeroLC23, DBLP:conf/chi/MirowskiMPE23}.}

\subsubsection{Adopting \toolname{} in Writing Tools}

\rev{\toolname{} is currently a standalone writing tool. 
However, the source code for the tool is available on {Github} (\textcolor{cyan}{\url{https://github.com/tonmoycsedu/HaLLMark/}}), allowing others to build upon it or modify it to their needs.
For example, with appropriate modification to our codebase, \toolname{} can be turned into a plugin.
Writers can then install \toolname{} in their favorite writing tool and track provenance information. Further, if implemented as a desktop application, \toolname{} can track provenance information across multiple tools (e.g., Microsoft Word, Overleaf, and Google Docs).
Alternatively, researchers and organizations can take inspiration from the design of \toolname{} and decide to build their own tracking interface.
For example, Microsoft Word already integrates LLMs into its writing interface.
Designing a tracking interface should be relatively straightforward. 
Finally, many visualization systems now store data using structured languages (e.g., Vega-Lite~\cite{DBLP:journals/tvcg/SatyanarayanMWH17}). One way to make the task of tracking provenance information platform-independent is to focus on storing provenance
information in a JSON format and then allowing copy and paste actions for the data across tools and devices.  } 

\subsubsection{\toolname{} as a Reading Tool}

\toolname{} is primarily focused on helping writers to ensure transparency and accountability in their creative content.
\rev{However, we believe that the provenance information collected by our approach could be shared in several ways with the recipients of written artifacts. For example, since \toolname{} is a web-based tool, writers could share a URL to the interactive document produced from \toolname{} as a supplement to the peer reviewers or publishers, who can then verify whether the artifact conforms to their writing policies using \toolname{}. 
Alternatively, \toolname{} could have a new feature that would create a static report of AI assistance (text highlighting on the artifact, summary statistics, and a list of prompts with the static timeline as an overview).
Writers could then share the report as a supplement to the actual artifact with the publishers or peer-reviewers. Finally, if our approach is able to store the data in a structured format (as discussed above), writers can store the provenance information and then share it with readers, who can then use their favorite reading tool to verify the contribution from the AI.
In all cases, publishers can decide how they want to share the provenance information with the larger audience.}





\subsection{Limitations}

\rev{\toolname{} is so far only a technology probe and is not designed for production use.
This means that it is limited in terms of the scale and scope of the documents and writing tasks it can produce.
While we did not perform any specific stress or performance testing in our evaluation, we believe that the current \toolname{} prototype implementation easily scales to a few thousand words.
While not evaluated, in theory, the visualizations in \toolname{} are also scalable to any number of prompts.
As mentioned in Section~\ref{sec:viz_prov}, there is a minimum threshold for the rectangle width (5px). 
When the rectangle width reaches the minimum, we dynamically increase the width of the SVG instead of decreasing the rectangle width and provide a horizontal scrollbar to see the extended contents. Prior visualization systems have adopted similar methods to make representations scalable~\cite{10292608}. 
Similarly, as policies evolve, we will likely see more constraints and dimensions, which could be encoded in the timeline by adding new rows in the timeline. Nevertheless, we acknowledge that as stories become longer and more variables are added to the timeline, it might become a daunting task to make sense of the timeline. One way to scale the representation is to aggregate the rectangles in the timeline, a popular approach in visualization design~\cite{10292608, DBLP:journals/tvcg/ElmqvistF10}. Our future work will explore these solutions. 
We have also discussed some practical manifestations of the tool in Section~\ref{sec:implications}; for example, a practical use case would be to implement the tool as a plugin that can be integrated with existing writing software, such as Grammarly.} 

Our study in this paper was limited to creative writers, and their experiences may differ from the general population of all writers. 
For example, compared to a fiction author, an academic or a journalist must rely on verifiable facts and evidence. 
This may alter the dynamic for such writers when using an LLM, as LLMs can generate hallucinations~\cite{ji2023survey}.
Further study is needed for these settings; they are beyond the scope of our present paper.

\subsection{Ethical Concerns of LLM based Co-Writing}

It could well be argued that the central argument of this paper---that LLMs are here to stay, and that we should just learn how to best leverage them---is a technopositive, na{\"i}ve, and perhaps even actively harmful approach to the use of AI in human creativity, and that generative AI should be seen as dangerous technology that should be regulated or even banned.
However, we would argue that this is true of virtually any technology.
For example, photography was widely hailed as the end of painting but instead freed painters from the curse of realism~\cite{Epstein_2023}.
Instead, by harnessing these technologies as supertools~\cite{Shneiderman2022} in support of and subservient to---and not partners or collaborators with---human writers is precisely the approach that we should be taking.

In the end, LLMs are just tools, even if they are highly sophisticated ones. 
By focusing on conveying the provenance between LLMs and \rev{humans}, essentially making human verification and influence the gold standard, we can reinforce \rev{this notion~\cite{Shneiderman2022, 10.1145/3544548.3580652, 10.1145/3593013.3594087}}.
After all, while many of us would view the idea of spending hours reading stories that were generated by a soulless machine somewhat insulting, most would likely accept this when assured that the overarching control of the story belonged to an actual human writer.
People already accept computer-generated imagery (CGI) in today's movies as a matter of course---why would they not accept similar computer-generated prose, as long as it has been verified (and potentially edited) by the author?
The provenance mechanisms presented in this paper, where these prompts, edits, and influences are made explicit in the text itself, is one approach to conveying this interaction history between the writer and the LLM.
Cryptology concepts such as NFTs~\cite{bamakan2022patents}---or placing the entire edit history on a blockchain---may be used to protect the integrity of this history.

\rev{Findings from our evaluation clearly indicate that our writer participants are mainly interested in using AI to improve their own writing rather than producing more copy faster.}
\rev{This may not strictly be true of students in educational settings, where LLMs could be argued to do more for the aspiring writer than act as the equivalent of a mere calculator for mathematics education.}
For example, one sentiment that \rev{was expressed by our participants, in line with previous studies~\cite{DBLP:conf/chi/MirowskiMPE23, DBLP:conf/ACMdis/GeroLC22},} is that writer's block may now be a phenomenon of the past, as the AI can always be relied upon to generate many new and fresh ideas of how to continue a story.
While we should always be wary of bad \rev{(or overworked and stressed)} actors that are indeed primarily seeking the ability to generate acceptable copy with a minimum of effort, \rev{professional writers harbor pride in the craft of writing~\cite{DBLP:conf/chi/GeroLC23}, as is true among virtually all professionals}.

Naturally, there are other ethical considerations that we must consider when putting this technology into the hands of writers. 
For one thing, it is possible that in spite of the tool's design to support author agency (as evidenced by the ability to writer's ability to edit or modify transaction history), other actors in the publication industry might be compelled to use the tool to surveil AI use and enforce AI writing policies.
Such has been the case for some academic writing support tools, such as Turnitin~\cite{turnitin}, which is designed to empower student learners, but has drawn criticism for its potential to police rather than support students.
Additionally, given the \rev{2023 strike} between the WGA and SAG-AFTRA on the one side, and the Alliance of Motion Picture and Television Producers (AMPTP) on the other, we should ensure not crossing any picket lines by actively making these tools freely available on the internet.
In the case of \toolname{}, while we have made the tool open-sourced on Github, we have added an explicit statement of support for WGA and SAG-AFTRA on the website as well as licensing terms prohibiting the use of the tool to cross the picket line.

\section{Conclusion}

We have presented a technology probe on AI co-writing called \textsc{\toolname{}} that enables an effective form of Large Language Model prompting while storing the provenance of interaction between human writer and AI. 
Designed based on our review of generative AI guidelines by professional and research organizations, \toolname{} transparently stores the prompting and influences between the LLM and the writer using text highlighting and a visual timeline.
We have presented our findings from a qualitative study involving a group of writers using \toolname{} to write a short story or non-fiction article. 
We \rev{found that our writers valued the explicit representation of the AI's influence on their work}, but also that the prompting interface yields a smoother and more integrated workflow than the default ChatGPT interface.

Human-AI co-writing is a nascent area of research that is also fraught with controversy. 
Our work addresses both transparency and prompting for LLMs supporting this modality, but is by no means the final nor optimal approach to either of these open research problems. 
We hope that future work can build on our findings to derive better supertools that retain human agency and control of the output while leveraging the formidable power of modern foundation AI models.
In particular, we think that future research should focus on \rev{scaffolding prompts, improving provenance tracking, and adding non-repudiation} of textual outputs generated by both human writers and AI models.

\begin{acks}
    While this work deals with AI co-writing, none of it was written using an AI model such as GPT-4. 
    In other words, \toolname{} was actually not used in producing the copy in this paper.
    This work was partly supported by grant IIS-2211628 from the U.S.\ National Science Foundation.
    Any opinions, findings, and conclusions or recommendations expressed here are those of the authors and do not necessarily reflect the views of the funding agency.
\end{acks}

\bibliographystyle{ACM-Reference-Format}
\bibliography{hallmark}

\end{document}